\documentclass[lettersize,journal]{IEEEtran}
\usepackage{amsmath,amsfonts}
\usepackage{algorithmic}
\usepackage{algorithm}
\usepackage{array}
\usepackage[caption=false,font=normalsize,labelfont=sf,textfont=sf]{subfig}
\usepackage{textcomp}
\usepackage{stfloats}
\usepackage{url}
\usepackage{verbatim}
\usepackage{graphicx}
\usepackage{cite}
\usepackage{hyperref}
\usepackage{tabularx}
\usepackage{multirow}
\usepackage{amssymb}
\begin{document}

\title{Generative AI-driven Cross-layer Covert Communication: Fundamentals, Framework and Case Study}
\author{Tianhao Liu, Jiqiang Liu,~\IEEEmembership{Senior Member,~IEEE}, Tao Zhang*,~\IEEEmembership{Member,~IEEE}, Jian Wang, Jiacheng Wang, Jiawen Kang,~\IEEEmembership{Senior Member,~IEEE}, Dusit Niyato*,~\IEEEmembership{Fellow,~IEEE}, Shiwen Mao,~\IEEEmembership{Fellow,~IEEE}.
        \thanks{Manuscript received XX XX, XXXX; revised XX XX, XXXX.This work is supported by the XXXXX.(Corresponding author: Tao Zhang, Dusit Niyato)}
        \thanks{T. Liu, J. Liu, T. Zhang and J. Wang are with the School of Cyberspace Science and Technology, Beijing Jiaotong University, Beijing 100044, China, and also with the Beijing Key Laboratory of Security and Privacy in Intelligent Transportation. Beijing Jiaotong University, Beijing 100045, China. (e-mail: leolth@bjtu.edu.cn; jqliu@bjtu.edu.cn; taozh@bjtu.edu.cn; wangjian@bjtu.edu.cn).} 
        \thanks{J. Wang, and D. Niyato are with the School of Computer Science and Engineering, Nanyang Technological University, Singapore 639798, Sinagapore (e-mail: jiacheng.wang@e.ntu.edu.sg;  dniyato@ntu.edu.sg). }
        \thanks{J. Kang is with the School of Automation, Guangdong University of Technology, and Key Laboratory of Intelligent Information Processing and System Integration of IoT, Ministry of Education, Guangzhou 510006, China, and also with Guangdong-HongKong-Macao Joint Laboratory for Smart Discrete Manufacturing, Guangzhou 510006, China (e-mail: kavinkang@gdut.edu.cn).}
        \thanks{S. Mao is with the Department of Electrical and Computer Engineering, Auburn University, Auburn, AL 36849-5201 USA. (e-mail: smao@ieee.org).}
        }

\markboth{IEEE Communications Magazine,~Vol.~XX, No.~XX, XX~XXXX}%
{Shell \MakeLowercase{\textit{et al.}}: A Sample Article Using IEEEtran.cls for IEEE Journals}


\maketitle
\begin{abstract}
        Ensuring end-to-end cross-layer communication security in military networks by selecting covert schemes between nodes is a key solution for military communication security. 
        With the development of communication technology, covert communication has expanded from the physical layer to the network and application layers, utilizing methods such as artificial noise, private networks, and semantic coding to transmit secret messages.
        However, as adversaries continuously eavesdrop on specific communication channels, the accumulation of sufficient data may reveal underlying patterns that influence concealment, and establishing a cross-layer covert communication mechanism emerges as an effective strategy to mitigate these regulatory challenges.
        In this article, we first survey the communication security solution based on covert communication, specifically targeting three typical scenarios: device-to-device, private network communication, and public network communication, and analyze their application scopes. 
        Furthermore, we propose an end-to-end cross-layer covert communication scheme driven by Generative Artificial Intelligence (GenAI), highlighting challenges and their solutions. Additionally, a case study is conducted using diffusion reinforcement learning to sovle cloud edge internet of things cross-layer secure communication.
        
\end{abstract}


\section{Introduction}
Military communications rely on a variety of methods, such as text messaging, satellite systems, and tactical signaling, to ensure seamless coordination and information flow in various operational domains\cite{jeonMilitaryNonterrestrialNetworks2024}. However, the growing complexity and interconnectedness of these channels introduce vulnerabilities, highlighting the need for more integrated and dynamic security solutions\cite{10608156}. These diverse channels are essential for maintaining operational effectiveness across various domains. However, they are inherently vulnerable to interception and information leakage, posing significant risks to security and mission integrity. The protection of these communication methods through advanced security protocols and countermeasures is crucial to protecting sensitive information and ensuring operational reliability.\par
Covert communication, also referred to as low detection probability communication, has emerged as a vital approach to improve communication security. It ensures reliable transmission for authorized users while preventing unauthorized parties from detecting any communication activities. This capability becomes particularly critical in military operations, where adversaries employ advanced eavesdropping and surveillance techniques. While encryption can secure the content of messages, it often relies on standardized protocols that cannot fully prevent adversaries from improving their decoding methods. Once these methods are enhanced, the risk of intercepting and leaking information increases significantly.
Covert communication can be categorized into three main types: physical layer, network layer, and application layer. 
\begin{itemize}
        \item \textbf{Physical layer:} 
        This involves techniques such as artificial noise generation and controllable coding to prevent adversaries from identifying modified communication data\cite{10440606}.
        \item \textbf{Network layer:}
        Private networks, often leveraging relay nodes such as Unmanned Aerial Vehicles (UAVs), adjust transmission energy levels and positions to construct secure channels and avoid adversarial detection. 
        \item \textbf{Application layer:}
        Information is embedded into the semantic redundancy of application data, such as video games or multimedia streaming\cite{sunTelepathMinecraftbasedCovert2023}, to achieve indirect communication without raising suspicion.

\end{itemize}

In practice, covert communications inevitably alter certain data characteristics, providing potential entry points for regulatory detection. 
\begin{itemize}
        \item \textbf{Active wardens:} Active wardens systematically collect and analyze data to progressively identify patterns of covert transmissions, focusing on specific channels through methods such as examining physical-layer signals, monitoring energy levels at the network layer, and modeling user behavior at the application layer\cite{9151255}.
        \item \textbf{Limited capacity:} Covert communication is implemented using methods such as artificial noise and lowering energy levels. Although these means enhance concealment, they usually significantly reduce transmission capacity, thereby affecting communication efficiency.
\end{itemize}
Nonetheless, the inherent challenge of effectively coordinating surveillance across different layers creates opportunities for cross-layer covert communication\cite{10793113}. Compared to single-layer approaches, a cross-layer strategy can maintain the same level of concealment while leveraging multiple channels to enhance transmission efficiency. 
\par
\par
In cross-layer covert communication, GenAI demonstrate significant advantages. By representing samples with latent vectors and learning their distributions, GenAI can generate realistic content and synthesize training sets, for example, simulating various complex network environments, device mobility patterns, and warden monitoring strategies.
Additionally, GenAI can be combined with decision making algorithms such as Deep Reinforcement Learning (DRL)\cite{Wang2022DiffusionPA} to construct optimal channel selection schemes, fully leveraging the unique advantages of covert communication at each layer. This not only improves the overall efficiency and concealment of the systems but also improves its adaptability and security in complex network environments, ensuring the continuity and reliability of military communication. The main contributions of this article are as follows:\par
\begin{itemize}
        \item \textbf{We provide a comprehensive tutorial on covert communication}, introducing fundamental concepts and characteristics at different layers, and demonstrating how the GenAI model enhances both efficiency and concealment, leading to more secure and robust communication strategies.
        \item \textbf{We propose a GenAI-driven cross-layer covert communication framework}, which enables flexible and efficient communication across multiple layers, improving both concealment and transmission performance in complex network environments.
        \item \textbf{We present a case study on CE-IoT using a diffusion reinforcement learning-driven method}, which shows how GenAI effectively tackles real-world challenges. This is the first work, to our knowledge, that leverages a diffusion-empowered reinforcement learning algorithm to optimize cross-layer covert communication, offering improved adaptability and security for IoT systems.
\end{itemize}
\section{Covert Communication Models: Basics, Current Progress, and Applications}
In this section, we comprehensively overview the different layers of covert communication, including the basics, current progress of different layers, and their applications.
\begin{figure*}[htbp]
\centering
\includegraphics[width=1\textwidth]{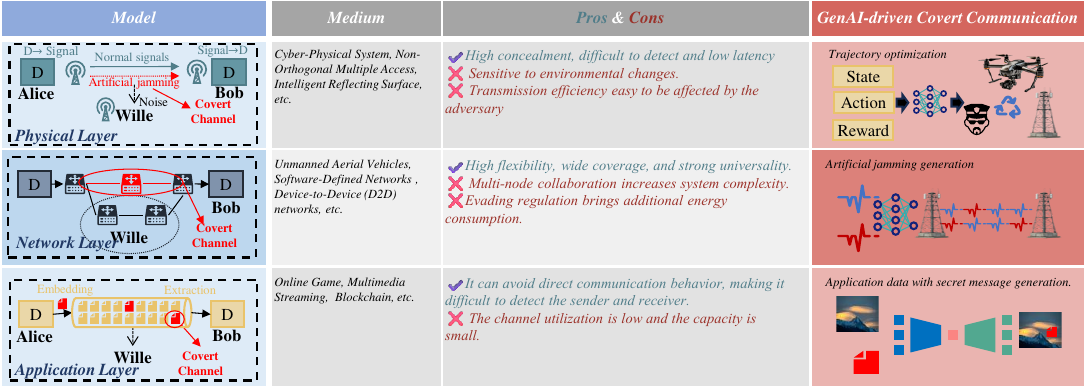}
\caption{The characteristics of covert communication schemes in different layers. }
\label{fig:summary}
\end{figure*}
\subsection{Basics}
\subsubsection{Covert Communication Definition} 
Often it is thought that the use of encryption is sufficient to secure communication. However, encryption only prevents unauthorized parties from decoding the communication. In many cases the simple existence of communication or changes in communication patterns, such as an increased message frequency, are enough to raise suspicion and reveal the onset of events. Covert channels aim to hide the very existence of the communication. Typically, covert channels use means of communication not normally intended to be used for communication, making them quite elusive.\par 
The typical model of covert communication can be divided into two stages: transmission by Alice and detection by Willie. 
Alice transmits her secret message $x[i]$ with a probability of $ p_1 $ to avoid detection by Willie and to improve her transmission performance. With the complementary probability $p_0$, Alice remains silent. Researchers commonly assume $p_1 = p_0 = 0.5$ to minimize Willie's detection probability. Conversely, Willie endeavors to accurately determine whether Alice is transmitting.
\subsubsection{Covert Communication in Different layers}
With the advancement of covert communication, it has evolved from focusing on Physical Layer Security (PLS) to encompassing network\cite{10093965} and application layer\cite{ivSecurityFoundationsApplicationbased2022}. The architectures and advantages and disadvantages of different layer are illustrated in Fig. \ref{fig:summary}.\cite{10440606} elaborates on the details of physical layer covert, \cite{10090449} delves into the specifics of network layer, and \cite{ivSecurityFoundationsApplicationbased2022} provides an in-depth discussion on application layer.
Covert communication schemes at different layers exhibit distinct characteristics. 
The physical layer transmits messages by adding artificial noise and controllable random variables to the signal. The network layer can evade adversaries by forming private networks, while the application layer uses semantic redundancy in the public network application data for indirect communication.
Next, we introduce the above three common covert communication models.
\subsection{Physical Layer Covert Communication}
Physical layer covert communication is designed to send information by physical signal without being discovered by adversary, it ensures confidentiality by altering the physical properties of the signal or exploiting the randomness of the channel. As Fig. \ref{fig:summary} shows, Alice embeds artificial noise into normal signals, making them receivable by Bob as secret information while appearing as noise to Willie. Common mediums include, but are not limited to Non-Orthogonal Multiple Access (NOMA), Intelligent Reflecting Surfaces (IRS)\cite{10793113} and so on. Based on these characteristics, physical layer covert communication exhibits high concealment and anti-jamming capabilities, while enabling uninterrupted and continuous transmission of secret messages.
The physical layer not only hides covert signals within normal communication but also provides a crucial foundation for cross-layer strategies. Through precise control of signal parameters at the physical layer, higher-layer protocols can seamlessly integrate covert channels with minimal overhead, thus enhancing both concealment and latency performance while evading detection mechanisms.
\subsection{Network Layer Covert Communication}
Network layer covert communication utilizes meticulously designed relay nodes and communication paths to achieve communication in private networks that are invisible to the warden.
As Fig. \ref{fig:summary} shows, Alice and Bob can communicate through relay nodes, such as UAVs, to avoid detection by Willie. 
In a network environment with multiple nodes, Alice and Bob use special nodes to construct a dedicated network, ensuring that data transmission does not pass through Willie. For example, users can use UAVs as relays and jammers, preventing ground-based Willies from detecting the communication\cite{10480325}. Common schemes, in addition to UAVs, include Software-Defined Networking (SDN), establishing adversary-inaccessible communication paths ,and creating covert communication links through Device-to-Device (D2D) networks.
In network layer covert communication, the optimization of physical layer signal transmission is achieved by constructing a dedicated network. By utilizing relay nodes and adjusting their positions, the impact of wardens on the channel can be reduced, thereby enhancing the covert nature and efficiency of the transmission.
\subsection{Application Layer Covert Communication}
Application layer covert communication is a technique that exploits the semantic redundancy of application data. By embedding secret information into the redundant fields of normal application data, it achieves covert message transmission.
As shown in Fig. \ref{fig:summary}, Alice designs a data embedding algorithm using the semantic redundancy of application data to embed secret messages into the application data. Bob extracts the secret messages by parsing the application data, while Willie, who does not know the extraction algorithm, cannot distinguish whether the application data has been embedded with secret information. Blockchain platforms, multimedia streaming websites, and even if online games can be used to embed messages\cite{ivSecurityFoundationsApplicationbased2022}. 
For example, users can re-encode online game data to embed secret messages without affecting normal transmission and parsing.\cite{sunTelepathMinecraftbasedCovert2023}.
Due to the influence of wardens, establishing communication between some nodes may be accompanied by significant communication noise or may only be possible at very low energy levels. Indirect communication using application layer data from public networks can effectively address these challenges.
\subsection{GenAI-driven covert communication}
GenAI analyzes large-scale data to identify patterns and predict traffic dynamics, improve resource allocation and reduce latency in communication networks. By leveraging historical data, GenAI reveals patterns in warden behavior, such as surveillance techniques and detection thresholds, allowing it to assess their impact on communication strategies and optimize them for maximum concealment. GenAI can also be used for concealment evaluation, where a legitimate user acts as the generator and the adversary as the discriminator to detect covert signals\cite{10440606}. Moreover, GenAI can generate controllable and indistinguishable content for covert communication carriers, further enhancing concealment\cite{10.1145/3703626}. 
For cross-layer covert communication, GenAI’s advanced content generation capabilities play a key role in overcoming the challenge of data insufficiency. By synthesizing realistic and diverse training datasets, GenAI can model various network scenarios, including adversarial behaviors and environmental changes, enabling a more comprehensive evaluation of the wardens influence across physical, network, and application layers.
\subsection{Lessons Learned}
The cases discussed above reveal both the advantages and challenges of single-layer covert communication in complex network environments, underlining the necessity of cross-layer schemes. From these insights, we derive the following key lessons:
\begin{itemize}
        \item Covert communication extending from the physical layer to the network layer and the application layer can enhance transmission capacity and improve concealment.
        \item Cross-layer communication increases network complexity, as the integration of multiple layers introduces new challenges in managing resources and ensuring seamless operation.
        \item GenAI enhances covert communication by optimizing resource allocation, predicting warden behavior, improving concealment through content generation, and addressing data insufficiency in cross-layer evaluations.
\end{itemize}
\section{GenAI-driven Cross-Layer Covert Communication Framework}
\begin{figure*}[htbp]
        \centering
        \includegraphics[width=0.95\textwidth]{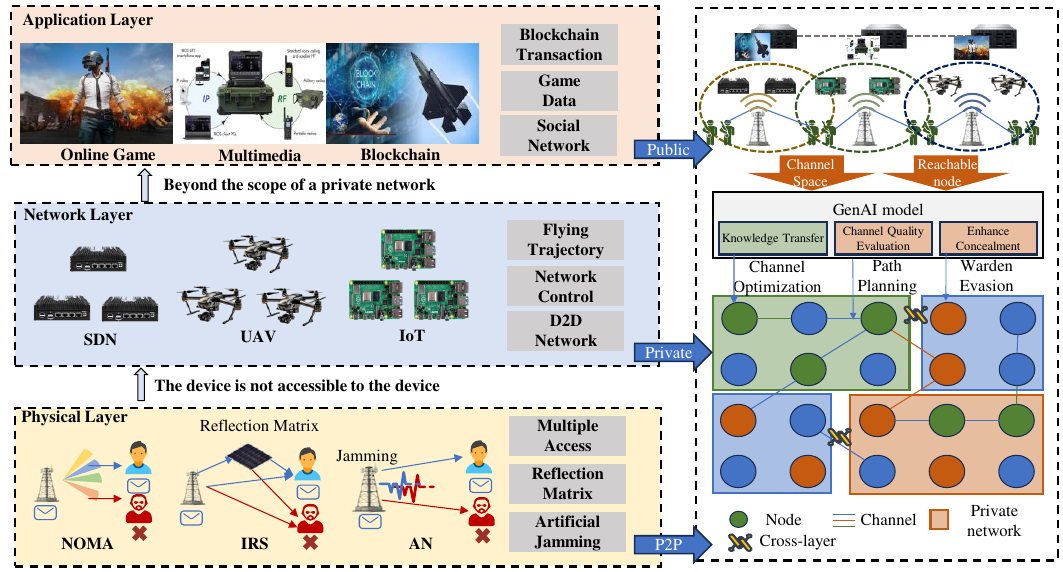}
        \caption{The framework of GenAI-driven cross-layer covert communication.}
        \label{fig:framework}
        \end{figure*}
In this section, we present a framework for GenAI-driven cross-layer covert communication, which integrates the physical, network, and application layer to achieve efficient and secure covert communication. Then, we introduce the major issues in cross-layer covert communication and discuss how GenAI can be used to address these challenges. Finally, we present the GenAI-driven framework.
\subsection{Framework Overview}
As shown in Fig. \ref{fig:framework}, the GenAI-driven cross-layer covert communication framework consists of three parts:
channel space, reachable node, and AI engine. 

\textbf{Channel Space:} 
The channel space is composed of various channels, including physical signals, information hiding, and semantic redundancy embedding between nodes. The diversity of channel types offers more options for covert carriers, allowing for the flexible selection of appropriate channels based on different types of monitors and environmental characteristics.
The channel space should ensure the reachability of covert communication between nodes, thereby guaranteeing smooth communication.

\textbf{Reachable Node:}
In cross-layer covert communication, nodes that can transmit information through covert communication are reachable nodes.
Some edge devices can receive physical layer signals and establish dedicated networks, serving as cross-layer relay nodes to achieve end-to-end covert communication, such as the control center of the UAV.
On the one hand, when point-to-point communication is under monitoring, the communication can be transferred to a dedicated network by searching for physical-to-network relay nodes among the reachable nodes, thereby evading the regulator. On the other hand, when the transmission target is not within the range of the dedicated network, indirect communication can be achieved through application layer covert channels using public networks.\par
\textbf{AI Engine:}
By introducing an AI engine at the application layer to drive cross-layer covert communication, an end-to-end covert communication framework is established. 
The AI engine takes the set of edges in the channel space and the set of reachable nodes as input, and constructs an end-to-end covert communication path based on concealment constraints, reachability constraints, and the goal of optimal communication efficiency.
GenAI models can provide powerful knowledge transfer capabilities. By evaluating the quality of all covert channels in the channel space and considering the impact of adversaries, GenAI models can develop optimal communication strategies that balance concealment and transmission efficiency.
\subsection{Major issues in Cross-Layer Covert Communication}
As shown in Fig. \ref{fig:framework}, the AI engine is designed to optimize channels, enhance information hiding capabilities, and counter warden.
Once the communication counterpart is determined, the AI engine constructs an end-to-end communication path by analyzing the channel space and reachable nodes. This process is constrained by warden influence and end-to-end connectivity requirements, with the primary objective of maximizing communication efficiency.
However, to achieve GenAI-driven end-to-end covert communication, the following issues need to be addressed:\par
\textbf{Channel Quality Assessment:} 
Due to the need for end-to-end communication data to be transmitted across different layers, it relies on various covert channels, posing significant challenges for channel quality assessment. For example, communication between two nodes can be achieved through a dedicated network or indirectly via a public network. In such cases, it is necessary to comprehensively consider factors such as concealment and communication efficiency to evaluate the advantages and disadvantages of both methods.\par
\textbf{Cross-layer warden influence:} 
In covert communication, the warden is typically a passive monitor, and their behavior directly impacts the concealment and reliability of different channels. For example, a warden sensitive to physical signals may pay more attention to physical layer covert communication while being less concerned with network layer covert communication. Therefore, the evaluation mechanism for the influence of warden on cross-layer channel needs to be reconsidered.
\subsection{Solutions From GDM}
Generative Diffusion Models (GDMs) are a powerful class of GenAI models based on probabilistic modeling. They can represent complex data distributions and generate high-quality samples\cite{10.1145/3703626}, allowing for the modeling and quantification of the concealment and efficiency of various channels, thereby providing accurate channel quality assessments. Furthermore, GDMs can be combined with reinforcement learning (RL) to transform the complex problem of cross-layer coordination into a Markov decision process. This combination generates numerous simulated state-action pairs to evaluate the performance of different channel combinations in various environments, thereby aiding decision-making and selecting the optimal channel paths.
\section{Case Study: Diffusion Empowered Cross-layer Covert Communication in CEIoT}
In this section, we perform a case study, presenting an diffusion empowered cross-layer covert communication method for Cloud Edge Internet of Things (CEIoT)\cite{10.1145/3555308}.
In recent years, the rapid proliferation of IoT devices, combined with the expanding capabilities of cloud and edge computing, has created vast opportunities for real-time data collection and analysis. Data are first sensed by IoT devices, then processed and transmitted by edge devices, and finally utilized in cloud servers, traversing multiple layers of communication. In contrast, data need to be protected from collection to use throughout the life cycle of privacy. \par
\begin{figure}
        \centering
        \includegraphics[width=0.5\textwidth]{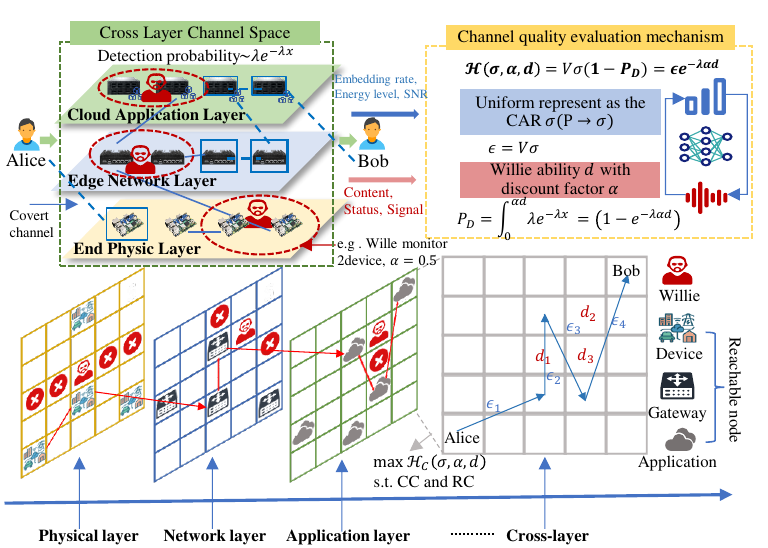}
        \caption{The illustration of cross-layer covert communication scheme in CEIoT, the influence of warden is detection ability $d$ and discount factor $\alpha$, which physically means warden's ability to detect data in the channel. Cross-layer multi-channel supervision dilutes its detection ability for each channel, so detection ability needs to be multiplied by $\alpha$.}
        \label{fig:case}
\end{figure}
To address these challenges, cross-layer covert communication has emerged as a promising solution to improve privacy protection in CEIoT systems.
There are diverse protocols, communication links, and device forms among the cloud, edge, and things end, and traditional security measures are unable to meet the diverse needs such as device heterogeneity and communication protocol diversity. Cross-layer covert communication can utilize different transmission protocols, networking methods, and redundant fields in existing protocol stacks to greatly enhance traffic anti-detection capabilities.\par
\subsection{System Model}
As fig. \ref{fig:case} shows, we use the cross-layer covert communication framework in CE-IoT for secure communication as a specific case. First, Alice wants to securely transmit the data to Bob. During this process, there are multiple Willies in the network who eavesdrop on the transmitted data to eavesdrop information and discard detected abnormal data. To ensure communication security, covert communication needs to be established to evade data monitoring. Next, Alice will identify reachable nodes and select available channels according to the network topology, thereby constructing a set of end-to-end covert communication paths. Furthermore, Alice uses the AI engine to evaluate channel quality from factors such as transmission capacity and Willie’s impact, and selects a set of channels that can reach Bob to construct the covert communication path.\par
To optimize the selection of covert communication paths, we propose a GenAI-driven channel quality evaluation mechanism. Specifically, we assume there are $N$ available channels in the channel space. Based on the channel parameters that meet the covert constraints, different channels evade surveillance by adjusting the signal-to-noise ratio (SNR) (physical layer), reducing energy levels (network layer), and adjusting data embedding rates (application layer). These methods will result in a decrease in transmission capacity. As shown in Fig. \ref{fig:case}, for a channel in channel space with the capacity of $V$, its covert channel capacity is represented by $V\sigma$, where $\sigma$ represents the Communication Availability Rate(CAR).
In addition, the actual transmission capacity of the covert channel depends not only on $\sigma$, but also on the fact that Willie’s eavesdropping can cause some of the transmitted data to be marked as abnormal and discarded. Therefore, it is necessary to consider data loss caused by Willie’s impact, which is quantified by the Detection Error Probability (DEP). The channel quality evaluation mechanism can assess the covert transmission capacity under Willie’s influence for each channel, aiming to maximize covert communication efficiency while satisfying covert constraints. 
\subsection{Problem Formulation}
We consider typically that covert user Alice wants to securely transmit data collected by an IoT device to the application user Bob (Fig. \ref{fig:case}), she needs to determine an end-to-end covert communication path composed of a set of covert channels $C$. $C$ needs to achieve the optimal transmission capacity while satisfying the reachability constraint (RC) and the Covert Constraint (CC). Specifically, the reachability constraint means that the channels in $C$ can form a path from Alice to Bob, and the covert constraint is expressed as all channels $i$ in $C$ should satisfy $P_{i}\gtrless_{D_1}^{D_0} \tau$ \cite{10090449}, where $D_1$ and $D_0$ represent two hypotheses: $D_1$: the channel is transmitting secret information, and $D_0$: the channel is not transmitting secret information. That is, under the threshold $\tau$, when the channel parameter $i$ (SNR, device energy level, application data embedding rate, etc.) settings are satisfied $D_0\le \tau\le D_1$, Willie will not consider this channel as a covert channel. Compared to $P_i$, the covert transmission capacity of the channel $i$ is $\epsilon_i=V_{i}\sigma_{i}$. Furthermore, due to Willie's detection of the channel, abnormal data will be discarded, which makes the transmission success rate affected by Willie's DEP $P_{D}$\cite{10793113}. Finally, we express the channel quality score evaluation formula as $\mathcal{H}(\sigma,P_{D})=V\sigma(1-P_{D})$, which means that channel quality is the channel covert transmission capacity multiplied by the transmission success rate. Then, the problem becomes finding the optimal combination of channels to maximize the channel quality score, which means this cross-layer covert communication path is optimal.
\subsection{Diffusion Reinforcement Learning-driven Cross-layer Covert Communication}
Currently, researchers use DRL models, such as Soft Actor-Critic (SAC)\cite{10675394}, to find optimal strategies. Furthermore, we integrate the diffusion process into traditional DRL to enhance its flexibility and exploration capabilities. We first detail our diffusion reinforcement learning-driven cross-layer channel selection. Notably, due to the lack of real cross-layer covert communication transmission data, we use brute-force computation as a substitute for the upper bound.\par
\begin{figure}[H]
        \centering
        \includegraphics[width=0.5\textwidth]{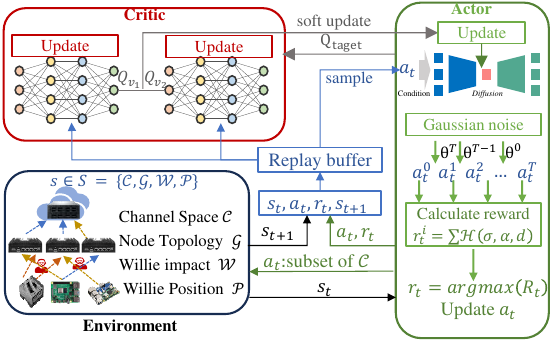}
        \caption{Diffusion empowered reinforcement learning model}
        \label{fig:dsac}
\end{figure}

As illustrated in Fig. \ref{fig:dsac}, we first define the state space based on the environmental context. The channel space represents the available covert communication channels, the node topology captures the fixed network structure, Willie’s capabilities model adversarial interference, and Willie’s position indicates the spatial locations of the adversary. 
The primary objective of the model is to identify the optimal channel selection strategy for cross-layer covert communication, guided by the channel quality evaluation mechanism. The actions involve selecting a set of covert channels that satisfy the Reachability Constraint (RC), evaluating their channel quality scores as feedback, and iteratively refining the model to learn an optimal channel selection strategy.
\par
Secondly, we use a conditional diffusion model to build a strategy generator. The diffusion model maintains a policy distribution that learns the optimal action based on the current state. During each training iteration, the model begins with Gaussian noise and undergoes a T-step denoising process, ultimately revealing the latent action space. This latent space follows a normal distribution, which aids in exploring strategies within complex or high-dimensional action spaces, thereby enhancing the agent's ability to identify the optimal action.\par
Finally, we initialize an actor network and two critic networks to evaluate the policy. At each time step, the agent obtains the current state from the environment and outputs the probability distribution of actions for that state through the actor network. The agent then samples an action from this distribution and executes it, receiving the new state and reward from the environment. Afterward, the two critic networks estimate the value of the state-action pair and the entropy of the action. Based on these estimates, the actor and critic networks are updated to optimize the diffusion process until the model selects the best action.
\subsection{Numerical Results}
\begin{figure*}
        \centering
        \includegraphics[width=0.88\textwidth]{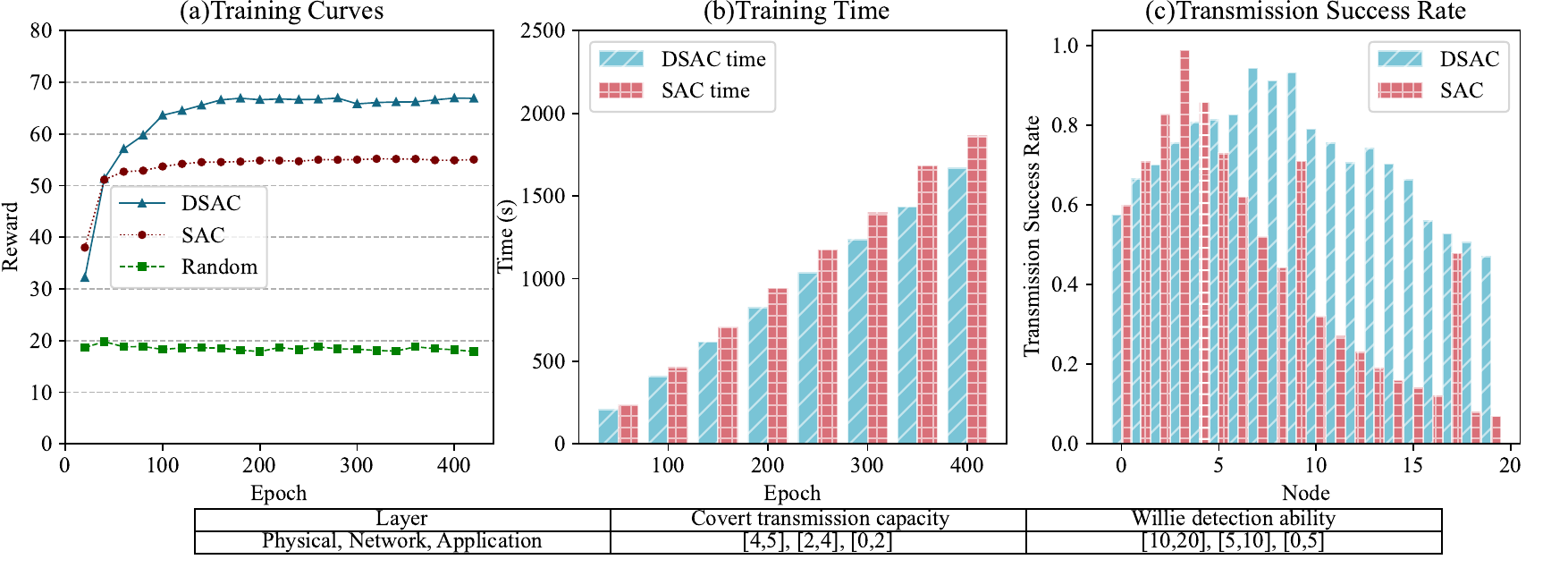}
        \caption{Numerical results of our diffusion-empowered SAC and conventional SAC}
        \label{fig:exp}
\end{figure*}

We conducted simulation experiments to demonstrate the effectiveness of the proposed scheme. We set the environment as a CEIoT network with 20 nodes, each with a channel space capacity of 9. The covert transmission capacity and Willie detection ability of each channel were randomly generated, with different value ranges set to simulate the different layers channel, the table in Fig. \ref{fig:exp} presents the parameter settings. 
We assume that Willie's detection ability is influenced by the distance between Willie and the initial node of the channel, with the discount factor being the reciprocal of the distance between them. The training objective is to learn the optimal channel selection strategy. The action involves selecting a set of channels from the channel space that satisfy RC. The reward is calculated through the channel quality evaluation mechanism.\par
Fig. \ref{fig:exp} depicts the training progress of the channel selection strategy generator based on Diffusion SAC (DSAC) and traditional SAC.
Firstly, as Fig. \ref{fig:exp}(a) shows, the rewards obtained by our method after convergence are significantly higher than those obtained by SAC. Moreover, on the current scale, our method essentially solves the optimal strategy, with optimal rewards ranging between 70 and 75 through brute-force computation. Secondly, as Fig. \ref{fig:exp}(b) shows, DSAC is better than SAC in training efficiency, which means diffusion model helps the strategy network to find the high-quality action distribution more quickly. Finally, after training (Fig. \ref{fig:exp}(c)), DSAC can select the channel with higher transmission accuracy, which proves that our method can reduce Willie's influence more effectively.
The superiority of the diffusion-empowered method is mainly attributed to diffusion exploration can enhance the flexibility of policy generation, preventing the model from getting stuck in suboptimal solutions\cite{duEnhancingDeepReinforcement2024}.
In contrast, our method enhances the transmission efficiency of cross-layer covert communication by leveraging the DSAC model and channel quality evaluation mechanism to select the optimal communication path.
\section{Conclusion}
This article presents a GenAI-driven cross-layer covert communication framework that addresses critical challenges in secure communications. By leveraging GenAI model, the proposed framework significantly improves operational security and data concealment in hostile environments. The implementation of diffusion empowered covert channel quality evaluation mechanism demonstrates the effectiveness of the framework in optimizing covert channel selection. Experimental results demonstrate that the incorporation of diffusion models can effectively enhance reinforcement learning performance, thus providing robust support for strategy generation in cross-layer covert communications. In future work, we will further explore the following aspects:\par 
\begin{itemize}
    \item Covert communication strategies under active supervisors, which account for their continuously evolving surveillance techniques, like dynamic detection methods and adaptive monitoring capabilities.
    \item Expanding Applications like Space-air-ground information networks and blockchain extension network, and conduct experiments in real-world scenarios to validate and refine the effectiveness of the proposed scheme.
\end{itemize}

\bibliographystyle{IEEEtran}
\bibliography{ref}
\end{document}